\documentclass[12pt,letterpaper]{article}
\usepackage[utf8]{inputenc}
\usepackage{amsmath}
\usepackage{amsfonts}
\usepackage{amssymb}
\usepackage{graphicx}
\usepackage{slashed}
\usepackage{textcomp}
\usepackage[left=3.00cm, right=3.00cm, top=3.00cm, bottom=3.00cm]{geometry}

\begin{document}
\title{Valence structures of light and strange mesons from the basis light-front quantization framework}

\author{Shaoyang Jia\footnote{E-mail: sjia@iastate.edu}~~and~James P. Vary\\ 
Department of Physics and Astronomy, Iowa State University,\\
Ames, Iowa 50011, USA}

\maketitle
\begin{abstract}
We apply the basis light-front quantization framework to solve for the structures of mesons with light and strange valence quarks. Our approach treats mesons as relativistic bound states with quarks confined in both the transverse direction and the light-front longitudinal direction. The spin-orbit interactions of these confined quarks are further specified by the Nambu--Jona-Lasinio model. We address the $\mathrm{U}(1)_{\mathrm{A}}$ axial anomaly by including the Kobayashi-Maskawa-'t~Hooft interaction regularized by our basis. We present the structures of the pion, the kaon, the eta meson, and the eta-prime meson in terms of their valence light-front wave functions obtained from the eigenvalue problem of our light-front Hamiltonian.
\end{abstract}

\section{Introduction to basis light-front quantizaton}
The light-front quantization framework~\cite{Brodsky:1997de} for quantum field theories specifies the commutation relation for field operators at equal light-front time $x^+ =x^0+ x^3$. Orthogonal to $x^+$, the light-front longitudinal coordinate is defined as $x^-=x^0-x^3$. The momenta conjugate to the light-front time and longitudinal coordinate are given by $p^\mp=p^0\mp p^3$, leaving the transverse momentum specified by $\overrightarrow{p}^\perp=(p^1,p^2)$. Because of the light-front quantization conditions, the Hamiltonian of the field theory is a well-defined hermitian operator in the Hilbert space covering all relevant Fock sectors. The resulting Hamiltonian dynamics allows non-relativistic many-body methods to be applied to problems in quantum field theories~\cite{Vary:2009gt}. 

Bound states structures are obtained from the wave functions as solutions of the light-front Schr\"{o}dinger equation 
\begin{equation}
\mathcal{H}_{\mathrm{eff}}\vert\Psi\rangle =M^2\vert\Psi\rangle.\label{eq:LF_Schrodinger}
\end{equation}
Here the effective Hamiltonian is defined as $\mathcal{H}_{\mathrm{eff}}=P^+\mathcal{P}^--\overrightarrow{P}^{\perp 2}$, where the light-front energy operator $\mathcal{P}^-$ is obtained from the Lagrangian using the Legendre transform~\cite{Brodsky:1997de}. While $\vert\Psi\rangle$ is the light-front state vector for the bound state with mass $M$ and conserved light-front $3$-momentum $(P^+,\overrightarrow{P}^\perp)$. Specifically for mesons, we truncate the meson state vector to the valence Fock sector such that~\cite{Jia:2018ary} 
\begin{align} & \big\vert\Psi(P^+,\overrightarrow{P}^\perp)\big\rangle
=\sum_{\mathrm{p,q},r,s}\int_{0}^{1}\dfrac{dx}{4\pi x(1-x)}\int\dfrac{d\overrightarrow{\kappa}^\perp}{(2\pi)^2}\,\psi_{\mathrm{pq}rs}(x,\overrightarrow{\kappa}^\perp)\nonumber\\
&
\quad \times b_{\mathrm{p}r}^\dagger(xP^+,\overrightarrow{\kappa}^\perp+x\overrightarrow{P}^\perp)\,d_{\mathrm{q}s}^\dagger((1-x)P^+,-\overrightarrow{\kappa}^\perp+(1-x)\overrightarrow{P}^\perp)|0\rangle,\label{eq:Psi_meson_qqbar}
\end{align}
with $b^\dagger_{\mathrm{p}r}$ and $d^\dagger_{\mathrm{q}s}$ being the creation operators for the quark and the anti-quark respectively. The valence light-front wave function (LFWF) $\psi_{\mathrm{pq}rs}(x,\overrightarrow{\kappa}^\perp)$ depends on the quark longitudinal momentum fraction $x$ and the relative transverse momentum $\overrightarrow{\kappa}^\perp$. The non-italic letters in the subscripts are flavor labels, leaving the italic subscripts specifying spins. 

Let us adopt the kinetic energy, the transverse confinement potential, and longitudinal confinement potential collectively in $H_0$ defined in Eq.~(10) of Ref.~\cite{Jia:2018ary} to construct the basis functions in the leading Fock sector of mesons~\cite{Li:2017mlw}. We then decompose the valence LFWF in basis functions:
\begin{equation}
\psi_{\mathrm{pq}rs}(x,\overrightarrow{\kappa}^\perp)=\sum_{n,m,l}\psi_{\mathrm{pq}}(n,m,l,r,s)\,\phi_{nm}\left(\overrightarrow{\kappa}^\perp/\sqrt{x(1-x)} \right)\,\chi_l(x).\label{eq:def_basis_expansion}
\end{equation}
Each term in Eq.~\eqref{eq:def_basis_expansion} is an eigenfunction of $H_0$.
The transverse basis function $\phi_{nm}$, the longitudinal basis function $\chi_l$, and the corresponding eigenvalue are identical to those in Refs.~\cite{Li:2017mlw,Jia:2018ary}. 

With the basis-diagonal Hamiltonian $H_0$ supplemented by an effective one-gluon exchange interaction, one can solve for the valence structures of heavy quakonia and $B_c$ mesons~\cite{Li:2017mlw,Tang:2018myz}. However for light and strange mesons, we supplement $H_0$ by chiral interactions instead of gauge interactions.

\section{The three-flavor Nambu--Jona-Lasinio interactions} 
When the explicit contributions to the structures of QCD bound states from gluons are omitted, the dynamics of quarks are specified by the global chiral symmetry, resulting in quark interactions given by the Nambu--Jona-Lasinio (NJL) model. Explicitly, the Lagrangian in the scalar channel of the color-singlet NJL model is given by~\cite{Klimt:1989pm}  
\begin{equation}
\mathcal{L}_{\mathrm{NJL},\mathrm{SU}(3)}^{(4)}=\overline{\psi}(i\slashed{\partial}-m)\psi  +G_{\pi}\sum_{i=0}^{8}\left[\left(\overline{\psi}\lambda^i\psi\right)^2+\left(\overline{\psi}i\gamma_5\lambda^i\psi\right)^2\right].\label{eq:def_lagrangian_NJL}
\end{equation}
Here the flavor decomposition of the fermion field is given by $\psi=(\mathrm{u},\mathrm{d},\mathrm{s})^{\mathrm{T}}$. Meanwhile, $\lambda^i$ are the flavor-space Gell-Mann matrices. 

%The Lagrangian in Eq.~\eqref{eq:def_lagrangian_NJL} is constructed to respect the global chiral symmetry, an approximate symmetry of QCD. In the reality of the strong interaction, this $\mathrm{SU}(3)_{\mathrm{A}}$ subgroup of the chiral symmetry is broken by both the nonvanishing quark mass and the dynamics of the strong force, both of which are simulated by terms in Eq.~\eqref{eq:def_lagrangian_NJL}. 
In order to account for the breaking of the $\mathrm{U}(1)_{\mathrm{A}}$ symmetry in the NJL model, the Kobayashi-Maskawa-'t\,Hooft determinant terms in the form of
\begin{equation}
\mathcal{L}_{\det}=G_{\mathrm{D}}\left[\det\,\overline{\psi}(1+\gamma_5)\psi+ \det\,\overline{\psi}(1-\gamma_5)\psi\right] \label{eq:def_KMtH_det}
\end{equation}
are supplemented to Eq.~\eqref{eq:def_lagrangian_NJL}. Specifically, determinants in Eq.~\eqref{eq:def_KMtH_det} are taken in the flavor space, resulting in 6-fermion interactions. Working in the meson valence Fock sector we need to reduce interactions in Eq.~\eqref{eq:def_KMtH_det} into those with up to $4$ fermions, during which the following divergent integral is encountered:
\begin{align}
\mathcal{R}(\alpha,\beta,b,L_{\mathrm{max}},N_{\mathrm{max}}) & \equiv \int_{0}^{+\infty}\dfrac{dp_1^+}{4\pi p^+_1}\int \dfrac{d\overrightarrow{p}^\perp_1}{(2\pi)^2} \int_{0}^{+\infty}\dfrac{dp_2^+}{4\pi p^+_2}\int \dfrac{d\overrightarrow{p}^\perp_2}{(2\pi)^2}\nonumber\\
&\quad \times (4\pi p^+_1)\delta(p_1^+-p_2^2) \,(2\pi)^2\delta\left(\overrightarrow{p}^\perp_1-\overrightarrow{p}^\perp_2\right)\big\vert_{(\mathrm{basis\text{-}regularized})}.\label{eq:def_basis_reg}
\end{align}

To regularized Eq.~\eqref{eq:def_basis_reg}, recall that in the practice of solving for the meson LFWF numerically, the following basis cut-offs are implemented~\cite{Jia:2018ary}:
\begin{equation}
0\leq l\leq L_{\mathrm{max}}, \quad 0\leq n\leq N_{\mathrm{max}} ,\quad \mathrm{and}\quad -M_{\mathrm{max}}\leq m\leq M_{\mathrm{max}}. \label{eq:basis_truncations}
\end{equation}
After expanding the delta functions in terms of basis functions within the basis cutoffs defined by Eq.~\eqref{eq:basis_truncations}, Eq.~\eqref{eq:def_basis_reg} becomes finite. Further, because the integrals over longitudinal momentum fractions carry no dimension while the divergence does, the limit of $L_{\mathrm{max}}\rightarrow+\infty$ can be safely reached. Consequently, we adopt 
\begin{equation} \lim\limits_{L_{\mathrm{max}}\rightarrow+\infty}\mathcal{R}(\alpha,\beta,b,L_{\mathrm{max}},N_{\mathrm{max}})\big\vert_{(\mathrm{basis\text{-}regularized})}=\dfrac{b^2}{8\pi^2}(N_{\mathrm{max}}+1),\label{eq:basis_reg_constant}
\end{equation}
as the basis-regularized version of Eq.~\eqref{eq:def_basis_reg}. 

\section{Results for the $\pi^0$, $\rho^0$, $\eta$, and $\eta'$}
After implementing the basis regularization, we obtain the following reduction of the $6$-fermion interaction to the $4$-fermion interaction:
\begin{equation}
\det \overline{\psi}(1\pm \gamma^5)\psi \rightarrow -2 \mathbf{m}_{\mathrm{a}}\mathcal{R}_{\mathrm{a}}\,\epsilon_{\mathrm{abc}}\epsilon_{\mathrm{agh}}\,\overline{\psi}_{\mathrm{b}}(1\pm\gamma^5) \psi_{\mathrm{g}}\,\overline{\psi}_{\mathrm{c}}(1\pm\gamma^5) \psi_{\mathrm{h}},
\end{equation}
where $\mathcal{R}_{\mathrm{a}}$ is the $\mathcal{R}(\alpha,\beta,L_{\mathrm{max}},N_{\mathrm{max}})$ for the quark of flavor ``$\mathrm{a}$''. We work in the $\mathrm{SU}(2)$ isospin symmetric limit, which prompts the decomposition of the valence LFWF into isospin symmetric and anti-symmetric components. The Hamiltonian then becomes block-diagonal for the isospin scalar states and the isospin vector states both in the flavor-nonet channel. 

Similar to Ref.~\cite{Jia:2018ary}, we ignore instantaneous interactions and self-energy corrections. The chiral interaction term of the Hamiltonian in the BLFQ-NJL model then becomes 
\begin{align}
H^{\mathrm{eff}}_{\mathrm{NJL}} & =\int dx^2\int d\overrightarrow{x}^\perp \bigg\{-G_{\pi}P^+\sum_{i=0}^{8}\left[\left(\overline{\psi} \gamma^i \psi \right)^2-\left(\overline{\psi}\gamma^5\lambda^i \psi \right)^2 \right]\nonumber\\
& \quad -G_{\mathrm{D}}P^+\left[\det\,\overline{\psi} \left(1+\gamma^5\right)\psi + \det\,\overline{\psi} \left(1-\gamma^5\right)\psi \right]\bigg\}.\label{eq:def_H_eff_NJL}
\end{align}
Equation~\eqref{eq:def_H_eff_NJL} supplements the basis-diagonal Hamiltonian $H_0$ to account for the chiral dynamics that is relevant for the light and the strange mesons. We then solve for the eigenstates from Eq.~\eqref{eq:LF_Schrodinger} using the combined Hamiltonian in our basis representation. For the isospin-vector mesons, we choose the strength of the determinant interaction such that $m_{\mathrm{s}}\mathcal{R}_\mathrm{s}G_{\mathrm{D}}=-G_\pi$. Together with the same $m_{\mathrm{l}}$ and $\kappa_{\mathrm{ll}}$ as those in Ref.~\cite{Jia:2018ary}, the LFWF of the neutral pion is identical to that of the charged pion in momentum space. 

To solve for the flavor wave function of the $\eta$ and $\eta'$, we apply the two-angle mixing scheme to calculate the decay constants and mixing angles~\cite{Feldmann:2002kz,Escribano:2005qq}. Also with the condition that $m_{\mathrm{s}}\mathcal{R}_\mathrm{s}G_{\mathrm{D}}=-G_\pi$, we obtain preliminary results for the masses, mixing angles, and decay constants of the $\eta$ and $\eta'$ mesons as listed Table~\ref{tab:eta_etap} using input parameters in Table~\ref{tab:input_eta_etap}. 
\begin{table}
	\centering
\caption{Input parameters for the BLFQ-NJL model of the $\eta$ and $\eta'$ mesons. }
\begin{tabular}{cccccccc}
\hline
$m_{\mathrm{l}}$ & $\kappa_{\mathrm{ll}}$ & $m_{\mathrm{s}}$ & $\kappa_{\mathrm{ss}}$ & $G_{\pi}$ & $N_{\mathrm{max}}$ & $M_{\mathrm{max}}$ & $L_{\mathrm{max}}$\\ 
\hline
300 MeV & 400 MeV & 450 MeV & 400 MeV & $1.7$ GeV$^{-2}$ & 8 & 2 & 8 \\ 
\hline
\end{tabular} 
\label{tab:input_eta_etap}
\end{table}
\begin{table}
	\centering
	\caption{Masses, mixing angles, and decay constants of the $\eta$ and $\eta'$ mesons.}
\begin{tabular}{ccccccc}
\hline
& $m_\eta$ (MeV) & $m_{\eta'}$ (MeV) & $\theta_8$ & $\theta_0$ & $f_8$ (MeV) & $f_0$ (MeV) \\ 
\hline
BLFQ & 554 & 939 & 4.23\textdegree & 16.4\textdegree & 211 & 89.4 \\ 
Ref.~\cite{Tanabashi:2018oca} & 547.86 & 957.8 & - & - & - & - \\ 
Ref.~\cite{Feldmann:1998vh} & - & - & -21.2\textdegree & -9.2\textdegree & $164.1$ & $152.3$ \\
\hline
\end{tabular} 
\label{tab:eta_etap}
\end{table}

\section{Conclusion}
We have expanded the basis light-front quantization framework with confinement and the Nambu--Jona-Lasinio interactions previously applied to the charged light and strange mesons~\cite{Jia:2018ary} to describe the valence structures of flavor-nonet mesons. Specifically, we have developed the regularization of divergences in the reduction of multi-fermion vertices consistent with our basis truncation. With a rough estimate of the model parameters, we obtained preliminary results for the masses, decay constants, and mixing angles of the $\eta$ and $\eta'$ mesons. With future searches in the model parameter space, we expect to find a consistent description of the valence structures for $\eta$, $\eta'$, $\omega$, and $\phi$ mesons within the BLFQ-NJL model. 

\section{Acknowledgments}
This work was supported by the Department of Energy under Grants No. DE-FG02-87ER40371 and No. DE-SC0018223 (SciDAC4/NUCLEI). 

\bibliographystyle{utphys}
\bibliography{SJ_BLFQ_bib}

\end{document}